\documentclass[twocolumn]{aastex631}

\usepackage{amsmath}
\usepackage{diagbox}
\usepackage{booktabs}
\usepackage{comment}
\usepackage[ruled,vlined]{algorithm2e}
\def\DM{\mathrm{DM}}

\def\pccm{\mathrm{pc}\,\mathrm{cm}^{-3}}
\newcommand{\D}{\mathrm{d}}

\newcommand{\EQ}[1] {Eq.~(\ref{#1})}
\newcommand{\FIG}[1] {Figure~\ref{#1}}

\usepackage{chngcntr}
\counterwithout{figure}{section}

\begin{document}

\title{Towards DM-free search for Fast Radio Bursts with Machine Learning -- I. An implementation on multibeam data}

\correspondingauthor{Rui Luo, Chen Wang}
\email{rui.luo@gzhu.edu.cn, chen.wang@csiro.au}

\author[0009-0005-5410-001X]{Yao Chen}
\affiliation{Department of Astronomy, School of Physics and Materials Science, Guangzhou University, Guangzhou 510006, China}

\author[0000-0002-4300-121X]{Rui Luo}
\affiliation{Department of Astronomy, School of Physics and Materials Science, Guangzhou University, Guangzhou 510006, China}

\author[0000-0002-3119-4763]{Chen Wang}
\affiliation{CSIRO Data61, 13 Garden Street, Eveleigh, NSW 2015, Australia}

\author[0000-0002-8744-3546]{Yong-Kun Zhang}
\affiliation{National Astronomical Observatories, Chinese Academy of Sciences, Beijing 100101, China}

\author[0009-0006-1500-441X]{Shiqian Zhao}
\affiliation{Department of Astronomy, School of Physics and Materials Science, Guangzhou University, Guangzhou 510006, China}

\author[0009-0003-8221-9611]{Chengbing Lyu}
\affiliation{Department of Astronomy, School of Physics and Materials Science, Guangzhou University, Guangzhou 510006, China}

\author[0009-0007-2280-1254]{ZePeng Zheng}
\affiliation{School of Physics and Astronomy, Sun Yat-sen University, Daxue Road, Zhuhai 519082, China}

\author[0009-0005-1848-0553]{Hai Lei}
\affiliation{Department of Astronomy, School of Physics and Materials Science, Guangzhou University, Guangzhou 510006, China}

\author[0000-0002-6423-6106]{DeJiang Zhou}
\affiliation{National Astronomical Observatories, Chinese Academy of Sciences, Beijing 100101, China}

\author[0000-0001-6651-7799]{Chenhui Niu}
\affiliation{Institute of Astrophysics, Central China Normal University, Wuhan 430079, China}

\author[0000-0002-9274-3092]{JinLin Han}
\affiliation{National Astronomical Observatories, Chinese Academy of Sciences, Beijing 100101, China}
\affiliation{School of Astronomy and Space Sciences, University of Chinese Academy of Sciences, Beijing 100049, China}

\author[0000-0003-1502-100X]{George Hobbs}
\affiliation{CSIRO Space and Astronomy, PO Box 76, Epping, NSW 1710, Australia}

\author[0000-0003-3010-7661]{Di Li}
\affiliation{New Cornerstone Science Laboratory, Department of Astronomy, Tsinghua University, Beijing 100084, China}

\author[0009-0008-7428-1665]{Chengwei Liang}
\affiliation{Department of Astronomy, School of Physics and Materials Science, Guangzhou University, Guangzhou 510006, China}

\author[0009-0005-3611-4143]{Siyi Tan}
\affiliation{Department of Astronomy, School of Physics and Materials Science, Guangzhou University, Guangzhou 510006, China}

\author[0009-0009-9276-3585]{Ting Tian}
\affiliation{Department of Astronomy, School of Physics and Materials Science, Guangzhou University, Guangzhou 510006, China}

\begin{abstract}

Searching for fleeting radio transients like fast radio bursts (FRBs) with wide-field radio telescopes has become a common challenge in data-intensive science. Conventional algorithms normally cost enormous time to seek candidates by finding the correct dispersion measures, of which the process is so-called dedispersion. Here we present a novel scheme to identify FRB signals from raw data without dedispersion using Machine Learning (ML). Under the data environment for multibeam receivers, we train the EfficientNet model and achieve both exceeding 92\% accuracy and precision in FRB recognition. We find that the searching efficiency can be significantly enhanced without the procedure of dedispersion compared with conventional softwares like \textsc{TransientX} and \textsc{presto}. Specifically, the impact of radio frequency interference (RFI) for single-beam and multibeam data has been investigated, and we find ML can naturally mitigate RFI under the multibeam environment. Finally, we validate the trained model on actual data from the current FRB surveys carried out by the Five-hundred-meter Aperture Spherical radio Telescope, which provides considerable potential for real implementation in the future.

\end{abstract}


\section{Introduction}

Fast Radio Bursts (FRBs) are extremely bright radio transients with typical durations from microsecond to millisecond scales. Observationally, FRBs exhibit a quadratic curve with prominent dispersion in the dynamic radio spectrum that records the flux of radio signal over frequency and time. Affected by diffused cold ionized plasma, in principle, the FRB signal at the high-frequency channel arrives earlier than that at the low-frequency channel. Since the time delay of pulse-like radio signals can be precisely measured, we would obtain the key information of cold plasma or ionized gas by fitting the equation as follows:
\begin{equation}
\Delta t = K_{\DM}\, \times \left[ \left(\frac{f_{\mathrm{low}}}{\mathrm{GHz}} \right)^{-2}-\left(\frac{f_{\mathrm{high}}}{\mathrm{GHz}} \right)^{-2} \right] \times \left( \frac{\DM}{\pccm} \right)\,,
\label{eq:delay_time} 
\end{equation}
where $K_{\DM}$ = 4.188808 ms is the dispersion constant, $\Delta t $ is the delay time, and $f$ is the observing frequency. The dispersion measure (DM) is defined quantitatively as an integrated column density of free or ionized electrons along the line of sight to the source.
\begin{equation}
   \DM = \int {n}_e \, \D l\,, 
\end{equation}
where ${n}_e$ is the electron density in units of $\rm cm^{-3}$ and $l$ is the distance traveled by electromagnetic waves in the interstellar medium, in units of parsec (pc).

Since the first FRB was serendipitously discovered in the data archive of pulsar survey \citep{Lorimer+07Sci}, there have been more than 800 FRB sources reported to date (see \textsc{blinkverse}\footnote{https://blinkverse.zero2x.org/}). FRB searching is typically done on high-time resolution radio astronomical data. How to conduct it effectively and accurately is still a big challenge. The most time-consuming part in conventional softwares is dedispersion, which requires finding the correct DM value to enhance the signal-to-noise ratio. Due to unknown DM, the traditional single-pulse searching method has to trial many different DM values in a broad range, which will cost huge amount of time, e.g., \textsc {TransientX} \citep{Men+24AA}, \textsc{presto} \citep{Ransom01PhDT} and \textsc{heimdall} \citep{Barsdell+12MN}. Once the signal-to-noise ratio is maximized through the optimal DM trial, a large number of single-pulse candidates would be produced for further analyses.

Nowadays, with the development of beamforming technology for telescope receivers, the telescope pointing system has realized monitoring or tracking at different adjacent positions simultaneously by using the coherent beams formed. The introduction of such receivers was motivated by the need to overcome the severely limited sky coverage of single-pixel feeds.  Multibeam observations enlarge the instantaneous field of view for blind surveys and therefore significantly increase the detection rate of transient events. 

\begin{figure}[hbpt]
\centering
\includegraphics[scale=0.8]{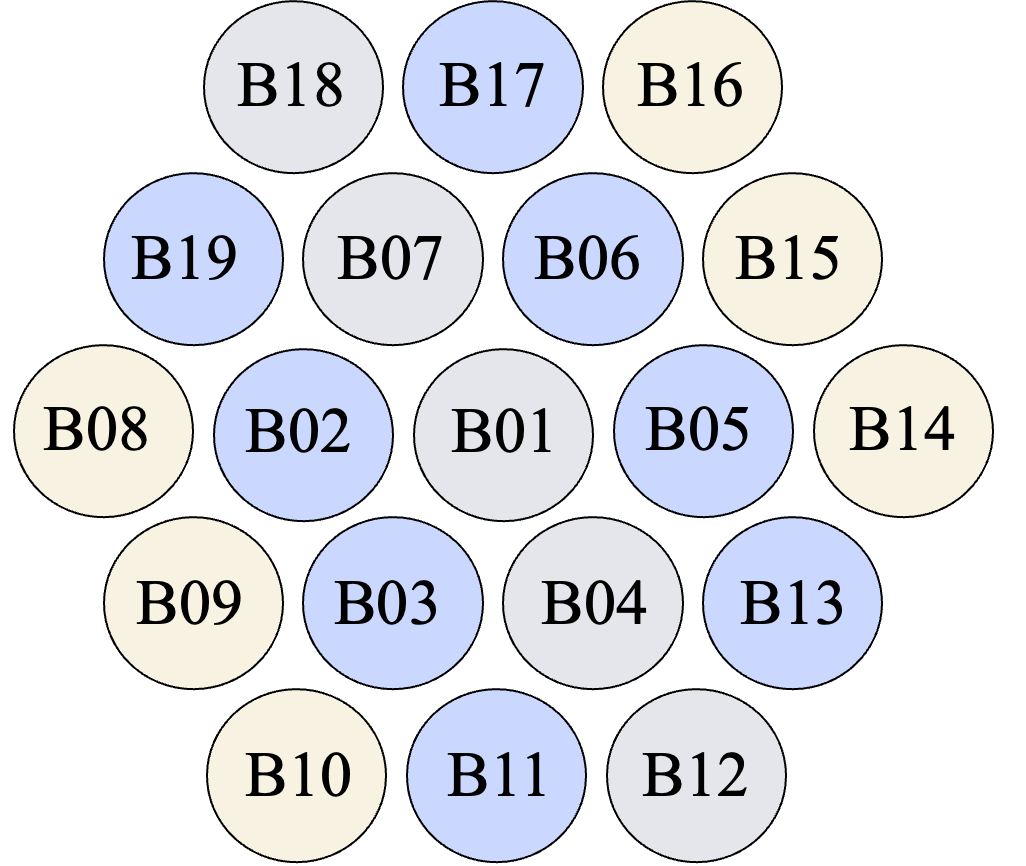}
\caption
{The layout of the FAST 19-beam receiver \citep{Jiang+20RAA}. The pointing system is centered on Beam 01, with Beams 02-19 surrounding the central beam.
\label{fig:19beam}}
\end{figure}

The Parkes 13-beam system laid the foundation for wide-field surveys such as HIPASS, the first blind HI map of the southern sky \citep{Barnes+2001MNRAS}, while the seven-beam ALFA receiver at Arecibo enabled high-sensitivity HI mapping (e.g., ALFALFA; \citealt{Giovanelli+2005AJ}) and greatly enhanced pulsar and transient search efficiency by surveying multiple lines of sight in parallel \citep{Cordes_2006}. FAST’s 19-beam array further combines unparalleled sensitivity with rapid sky coverage, enabling deep and wide surveys for HI, pulsars, and fast transients \citep{Nan+11IJMPD,Li+18IMMag,Jiang+20RAA}. The approximate distribution of the FAST 19-beam array is shown in Figure~\ref{fig:19beam}. More recently, next-generation wide-field receivers based on phased-array technology are being developed, such as the cryogenically cooled phased-array feed (CryoPAF) for the Parkes radio telescope \citep{Dunning+23}. Under the intensive data stream by multibeam or even phased-array receivers, conventional FRB searching algorithms relied on many DM trials would be challenged essentially.

Another challenge comes from radio frequency interference (RFI), which is unwanted signals produced by humans' electronic devices or space detectors. On the one hand, RFI can easily contaminate and even saturate the wideband receiver of a radio telescope \citep{Hobbs+2020PASA}. On the other hand, strong RFI can mask FRB signals, increase the number of false positives, and severely contaminate FRB surveys. 
For wide-field radio telescope, RFI with terrestrial origin is usually captured by all beams and exhibits similar or even identical characteristics in the dynamic spectra. But an FRB merely appears within one or several beams, for example, ``Perytons'' \citep{BSS+11ApJ, Petroff+2015MN} and the Lorimer Burst \citep{Lorimer+07Sci}. 

Regarding this apparent difference, it is realistic to extract the feature of FRBs compared with RFI from the multibeam data environment without dedispersion. Furthermore, the data collected by multibeam receiver has tremendously high transfer rate, the computing load arising from dedispersion for the data from each beam will be increased exponentially. These motivate us to think about an efficient, DM-free searching strategy that can be operated directly on multibeam data.

Machine Learning (ML) is worth deploying for data-intensive science, especially for high-time resolution radio observations. At present, ML has been widely used in searching for pulsars and FRBs. To sift pulsar candidates effectively, \cite{Zhu+14ApJ} developed a toolkit named ``PICS'' for pulsar candidate sorting. The tool has adopted many ML-related utilities, e.g., artificial neural network. About FRB searching, \cite{Connor+18AJ} applied a Deep Learning framework to classify single-pulse events of FRBs and pulsars, \cite{Zhang+18ApJ} trained the model based on the Convolutional Neural Network (CNN) and detected dozens of faint bursts from the repeating source FRB 121102. One of the most successful applications is, \cite{Agarwal+20MN} developed a Python package \textsc{FETCH} for classifying candidate signals in which Transfer Learning was used for model training. In addition, \citep{Zhang+20AA} has demonstrated that saliency map, which is an image feature extractor, can enhance the features of transient signals and make them stand out compared with RFI. Recently, \cite{Zhang+25ApJS} developed \textsc{DRAFTS}, a Python package based on deep learning that integrates object detection and binary classification techniques to accurately identify FRB.

In this paper, we propose an ML method for training and testing FRB detection. This method leverages multibeam data to mitigate RFIs and greatly reduces the computational cost of the traditional de-dispersion step. The paper is organized according to the following structure. In Section~\ref{sec:Method}, we introduce the methodology and the machine learning, in Section~\ref{sec:Dataset} describe the data, in Section~\ref{sec:Result} is the experimental results we obtained, and in Section~\ref{sec:Discussion} is the discussion. The relevant conclusions will be given in Section~\ref{sec:Conclusions}.

\begin{figure*}[htbp]
\includegraphics[width=0.98\textwidth]{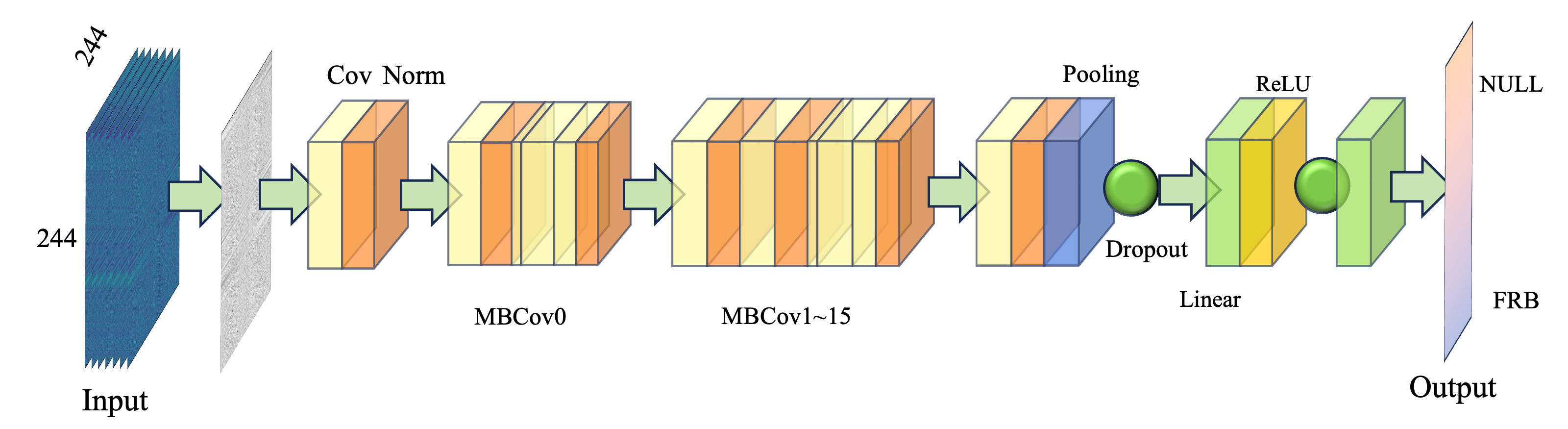}
\label{fig:model}
\caption{The architecture of the model used in this work. The input data is preprocessed to 224 pixels $\times$ 224 pixels $\times$ 7 channels, the EfficientNet model \citep{Tan+19arXiv} is used as the base model, the output feature of the fully connected layer of the model is modified to 256, and an additional Rectified Linear Unit (ReLU) activation function layer, Dropout layer, and a fully connected layer with an input feature number of 256 and an output feature number of 2 are added. Finally, the predicted output of the image classification is obtained.}
\end{figure*}

\section{An EfficientNet-based FRB searching model} \label{sec:Method}

We adopt the EfficientNet \citep{Tan+19arXiv} architecture as the base for our FRB detection algorithm. EfficientNet is a family of convolutional neural networks that can achieve high accuracy with fewer parameters through compound scaling of the network depth, width, and image size. It is suitable for high-throughput FRB detection by balancing the classification performance and execution efficiency. 

The architecture of the EfficientNet model we use is constructed in \FIG{sec:Method}. To accommodate diverse input data formats, we add a layer as an adapter before the first layer of the pretrained EfficientNet model. We set the dimensions of this new input layer to match the dimensions of the multibeam data from the telescope. We also add additional layers on top of the second-to-last layer of EfficientNet to make it output binary classification results. The additional layers include a ReLU layer, a dropout layer, a linear layer with 512 dimensions that produces 256 dimension output for a ReLU layer and a dropout layer followed by a two-dimensional output layer. Finally, we use the cross-entropy loss to fit training data and stochastic gradient descent (SGD) to train the model.

\section{Data and Training} \label{sec:Dataset}  

\subsection{Simulation setup and parameter space}

High-time resolution radio data typically records multiple aspects of information. We focus on parameters such as sampling time, signal frequency, signal strength, number of beams, relative positions between beams, and the Fourier transform of data from different channels. These also help to distinguish signals from astrophysical sources in environments with RFI.

Compared to actual data, simulated data have several unique advantages. First, it is easily accessible, unconstrained by observational conditions, and can generate high temporal resolution data whenever needed. Second, we can set up the parameters for simulation patterns flexibly, such as flux density, pulse width, DM and time of arrival at specific frequency. Furthermore, the inherent labeling of the simulated data provides an accurate standard for model training and performance evaluation.

We train the model on single-beam and multibeam datasets separately. To build the training datasets for the multibeam case, we need to create a realistic environment for the FAST 19-beam observation following its actual layout in the equatorial coordinates. The signals we simulate are all embedded in the 19-beam data sets.

\subsection{Simulation Data}
Here we use simulateSearch\footnote{\href{https://bitbucket.csiro.au/scm/psrsoft/simulatesearch.git}{https://bitbucket.csiro.au/scm/psrsoft/simulatesearch.git}} \citep{Luo+22MN} to build our training data sets, which include system noise, RFI, pulsars, FRBs, etc. In general, we specify the exact beam position of the FAST 19-beam receiver one by one (see \FIG{fig:19beam}), and simulate the data with 1-bit digitizer at different right ascensions and declinations. Each file has a duration of 400-second with a frequency range of 1 -- 1.5 GHz, and contains 100 signals that can be divided into 496 subintegrations (subint). As first defined by \cite{Hotan+2004PASA} and explained on the ATNF website\footnote{\url{https://www.atnf.csiro.au/research/pulsar/psrfits_definition/Psrfits.html}}, these sub-integrations are formed by integrating continuous time-series data over shorter time intervals.

For a given radio telescope, according to the radiometer equation, the flux density of its system noise can be described as follows. 
 \begin{equation}
S_{\nu,\mathrm{rms}} = \frac{T_{\mathrm{sys}}}{G \sqrt{N_\mathrm{pol}\Delta\nu\tau}}
\label{eq:noise}
\end{equation} 
where $S_{\nu,\mathrm{rms}}$ is the root-mean-square flux density of the noise, $ T_{\mathrm sys} $ is the system temperature, $G$ is the telescope gain, $N_{\mathrm pol} $ is the number of polarization channels, $\Delta\nu$ is the bandwidth of receiver and $\tau$ is the sampling time. To simulate the white noise for the FAST telescope, we adopt the specifications of the 19-beam receiver, as shown in Table \ref{tab:sys_params}.

\begin{table}[htbp]
    \centering
    \caption{\footnotesize Parameters of the FAST telescope system for simulation data.}
    \setlength{\tabcolsep}{12mm}{\begin{tabular}{lc}
        \hline
        \hline
        \textbf{Parameter} & \textbf{Value} \\ 
        \hline
        $G$\,(K/Jy) & 15   \\
        $T_{\rm sys}$\,(K)& 23  \\
        Freq. (GHz) & 1.0 -- 1.5   \\
        $N_{\rm chan}$ & 512 \\
        $t_\mathrm{samp}\,(\mu$s)  & 196.608   \\
        $N_{\rm bit}$ & 1 \\
        \hline
        \label{tab:sys_params}
    \end{tabular}}
\end{table}

\begin{figure*}[htbp]

\centering
\begin{minipage}{0.245\textwidth}
    \centering
    \includegraphics[width=\linewidth]{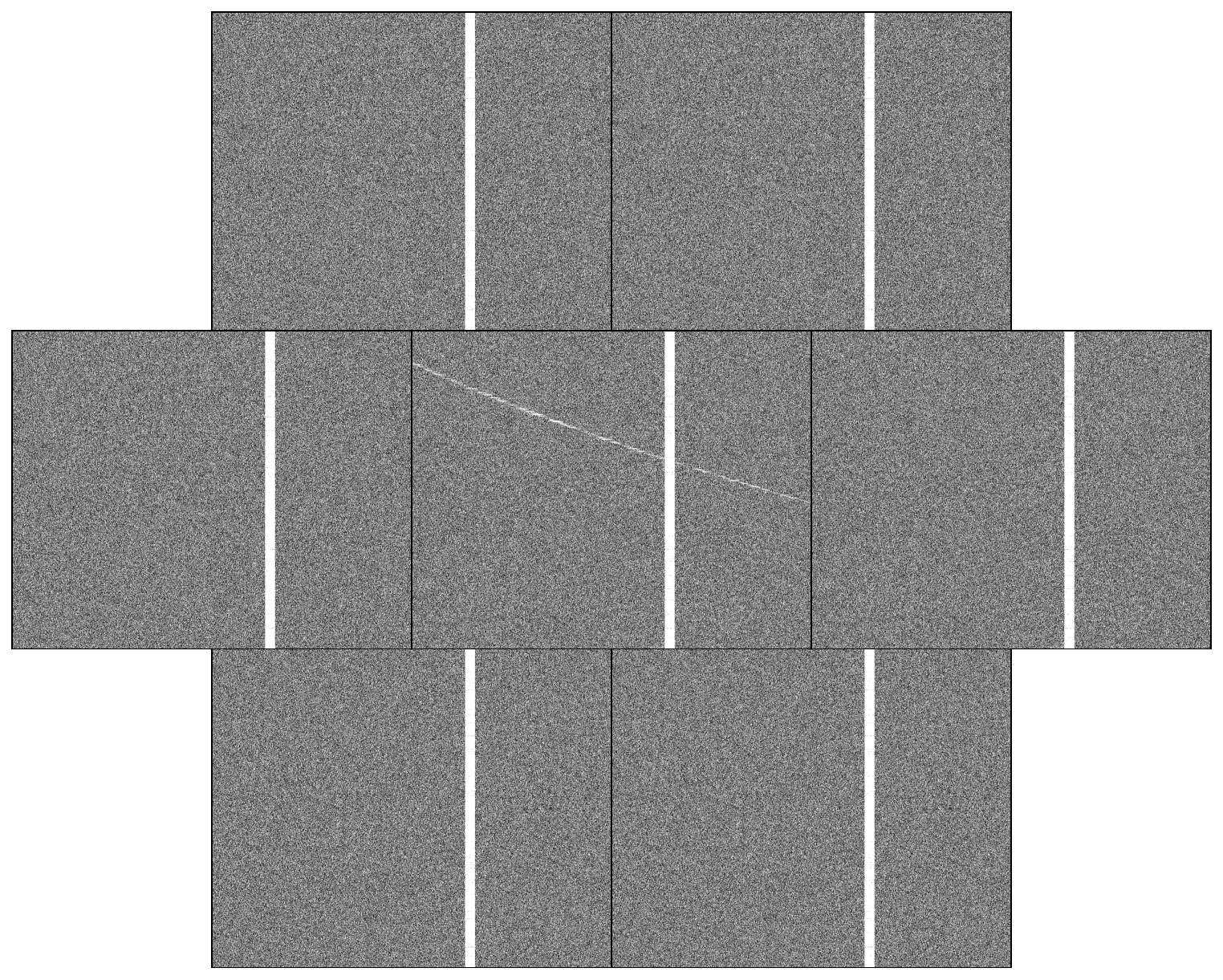}
\end{minipage}\hfill
\begin{minipage}{0.245\textwidth}
    \centering
    \includegraphics[width=\linewidth]{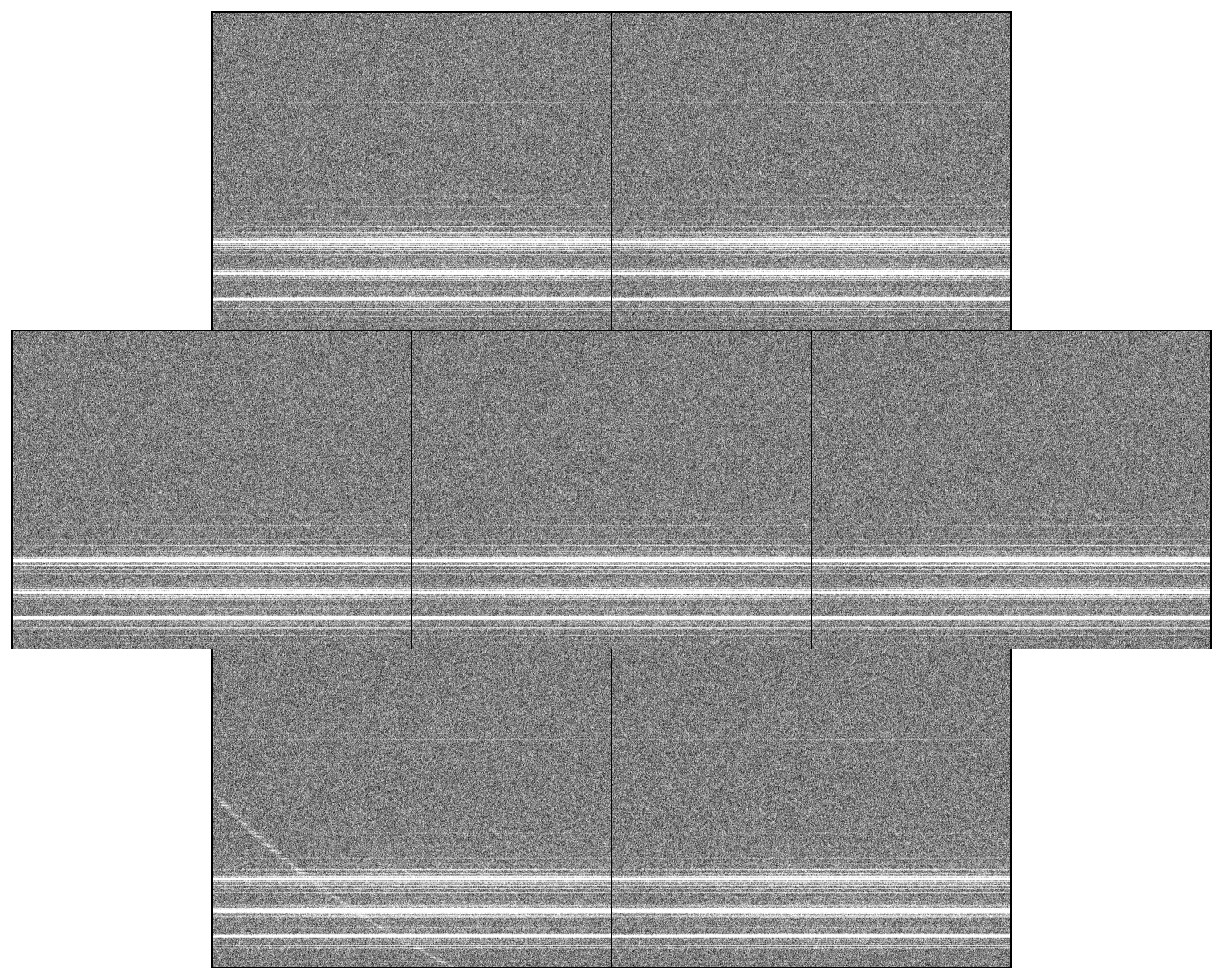}
\end{minipage}\hfill
\begin{minipage}{0.245\textwidth}
    \centering
    \includegraphics[width=\linewidth]{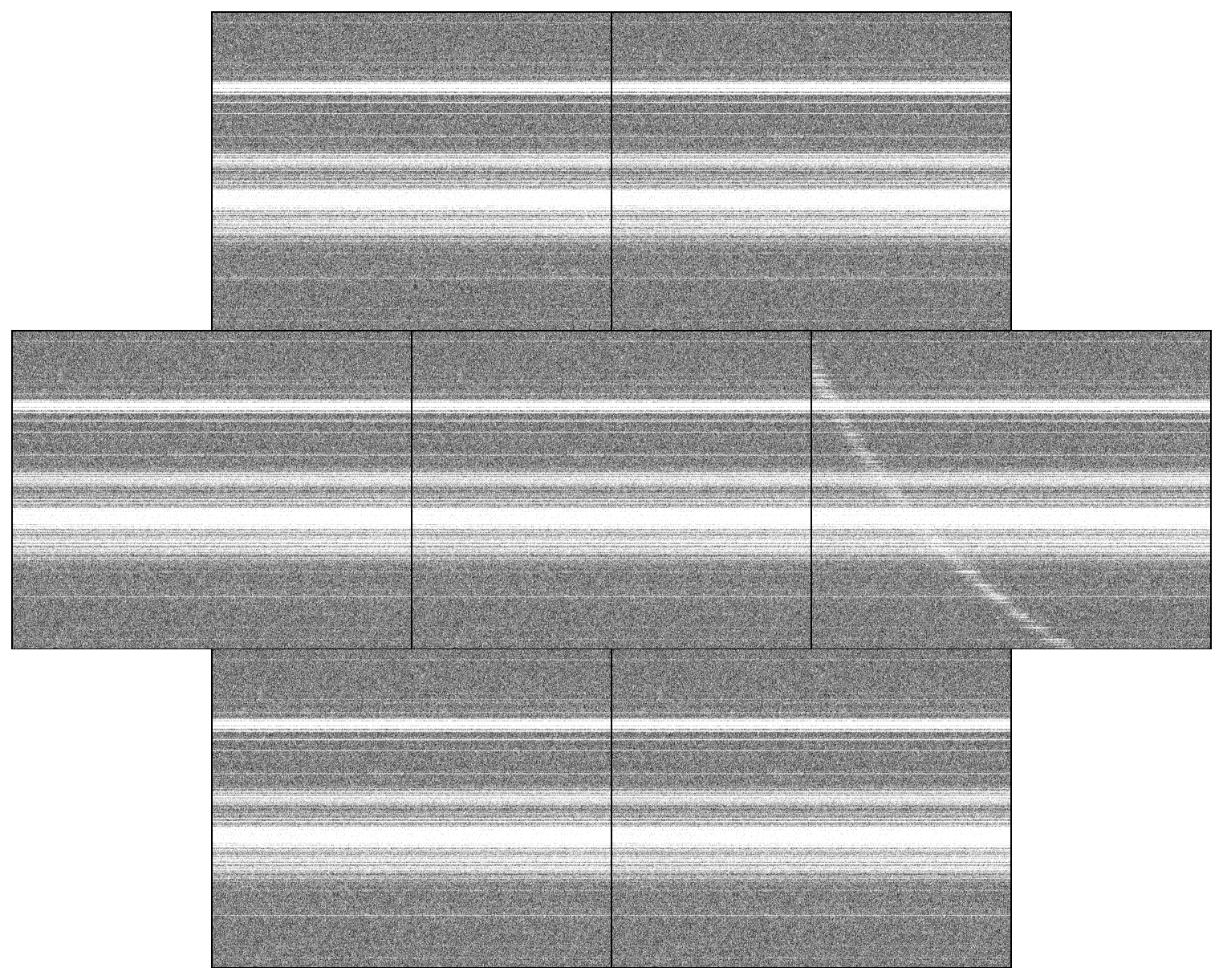}
\end{minipage}\hfill
\begin{minipage}{0.245\textwidth}
    \centering
    \includegraphics[width=\linewidth]{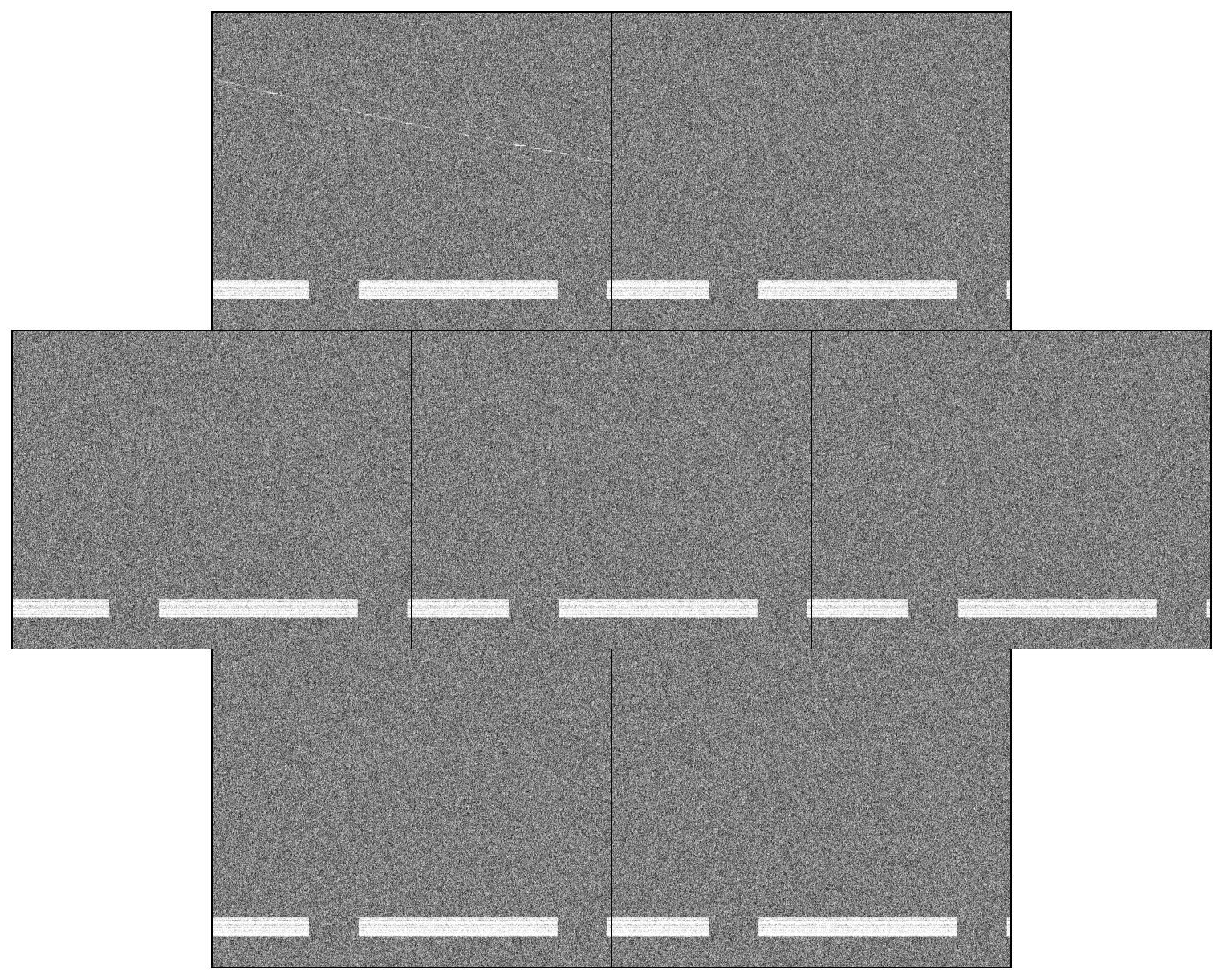}
\end{minipage}\hfill
 \vspace{0.5cm} 
\caption{Training samples of multibeam datasets. From left to right, the types of RFI are impulsive, point-to-point, satellite, and narrowband. In each subfigure, RFI is distributed across 7 beams, while FRB only appears randomly in one beam. In the panels of each sub-figure, the x-axis and y-axis represent time and frequency, respectively. Note that the samples in the multibeam dataset are formed by combining the central 7 beams into 7 channels used in \FIG{fig:model}.}
\label{fig:beam7_sample}
\end{figure*}

Under the routine for FRB simulation in \textsc{simulateSearch}, one needs to input the key parameters, i.e., arrival time, reference frequency, flux density, width, and DM. All the burst profiles are Gaussian-like determined by the peak flux and pulse width. In order to make the mock FRBs look realistic for a given telescope, as well as convenient for ML model training and validation, we simulate FRBs following these distributions: the sampled flux densities follow a power-law distribution from 0.0045 to 0.058\,Jy, and the sampled dispersion measures follow a uniform distribution from 110 to 1000\,$\pccm$. The pulse widths range from 0.001 to 0.02\,s, following a normal distribution with a mean of $\mu$ = – 2.5 and a standard deviation of $\sigma$ = 0.3.

To mimic the real blind search, the FRB signals are simulated in only one of the 19 beams, while RFI is placed in all beams. The RFI signals can be created realistically using \textsc{simulateSearch}, according to long-term monitoring in the real electromagnetic environment. For the FAST telescope, we simulate various types of RFI, including satellite RFI, narrowband RFI, impulsive RFI, and point-to-point RFI, which have been described in \cite{Luo+22MN}. 

In total, we simulate 988 data sets for ML model training, consisting of 152 with only FRBs, 228 with only RFI, and 608 with both RFI and FRBs. Each data set contained 19 files, which follow the FAST 19-beam. We divide all the data into two sets for different uses: 80\% for training and validation, and 20\% reserved for testing. The training-validation portion is then subdivided into training and validation sets using an 8:2 ratio. We combine the middle 7 beams from the same data set of 19 beams into three-dimensional data (time, frequency, and number of beams) as training data for the model in multibeam format.

\subsection{Data Pre-processing}  

We use simulation data to train the model. The data are converted to 8-bit format using \texttt{astropy} and \texttt{numpy} from \textsc{python}, then reconverted to 1-bit format using \texttt{unpackbits} from \textsc{python}. We further partition the data based on subint. Each subint is an input data sample of the model. If the input contains an FRB, it is assigned a positive label; otherwise a negative label. Whether an input contains an FRB follows the criterion given in Algorithm-\ref{alg:subint_label}.

\begin{algorithm}[htbp]
\caption{Identifying whether each subint is a positive sample or not} 
\label{alg:subint_label}
\KwIn{$t_{1250}$: The arrival time of signal at 1250\,MHz;
      $\Delta t_1$, $\Delta t_2$: Time offsets between 1250\,MHz and 1500/1000\,MHz calculated based on \EQ{eq:delay_time};
      $\{(t^{(i)}_{\text{start}}, t^{(i)}_{\text{end}})\}$: Subint time intervals}
\KwOut{Labels $\{label^{(i)}\}$ for each subint}

Compute $t_{1500} \leftarrow t_{1250} - \Delta t_1$\;
Compute $t_{1000} \leftarrow t_{1250} + \Delta t_2$\;

\For{$i$-th subint}{
    \If{$t_{1500} \le t^{(i)}_{\text{end}}$ \textbf{and} $t_{1000} \ge t^{(i)}_{\text{start}}$}{
        $label^{(i)} \leftarrow 1$\;  \tcp{Signal overlaps with this subint}
    }
    \Else{
        $label^{(i)} \leftarrow 0$\;  \tcp{No signal in this subint}
    }
}
\Return $\{label^{(i)}\}$\;

\end{algorithm}

For each input data, we apply bilateral filtering to smooth it as an image before applying the \emph{resize} function in the \texttt{skimage} package, in order to change its dimension to $224 \times 224$ and match the dimension of EfficientNet's input layer. For multibeam data, we stack them together to form a $224 \times 224 \times 7$ input. \FIG{fig:beam7_sample} shows four training samples of multibeam data after smoothing.

\subsection{Model Training}

We utilize the PyTorch library\footnote{\href{https://pytorch.org/get-started/locally}{https://pytorch.org}} to build our EfficientNet-based ML models. 
In practice, we use the cross-entropy loss to measure the difference between the predicted result and the label of the input. The stochastic gradient descent (SGD) optimizer is used to update model parameters. We set the batch size to 64 and the learning rate to 0.01\%, and the initial weight is 0.9. 
After each training epoch, the model is validated on the validation set. To avoid overfitting, we use Cosine Annealing to adjust the learning rate at each epoch and apply L2 regularization to the loss.

Our training experiments are run on two different computer nodes: One is equipped with a dual-circuit Intel Xeon Platinum 8358P processor (128 cores, 2.6 GHz) and 512 GB of RAM, mainly for data processing; The other one is equipped with two NVIDIA GeForce RTX 4090 graphics cards (24 GB of graphics memory per card), mainly for model training.

\section{Results} \label{sec:Result} 

\begin{figure}[htbp]
    \centering
    \begin{minipage}{0.40\textwidth}
        \centering
        \includegraphics[width=0.9\linewidth]{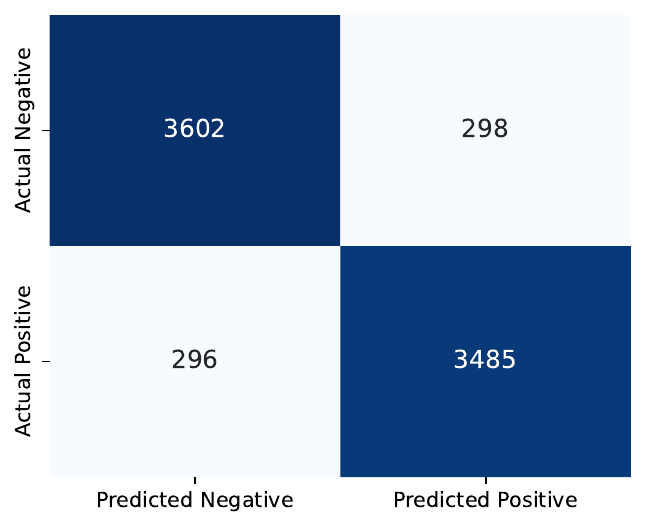}
        
    \end{minipage}\hfill
    \caption{Confusion matrices of the multibeam model. From left to right, the first row is TN, FN, and the second row is FP, TP. N or P denotes whether the model predicts the sample to be a positive sample or a negative sample. T or F indicates whether the model is correct or wrong.}
   
    \label{fig: matrix}
    
\end{figure}

To evaluate the performance of our model on test data, the performance metrics are defined below, including accuracy, precision, recall or true positive rate, F1 score, and false positive rate (FPR): 

\begin{equation}
    \text{Accuracy} = \frac{\text{TP} + \text{TN}}{\text{TP} + \text{FP} + \text{FN} + \text{TN}}\,,
\end{equation}

\begin{equation}
    \text{Precision} = \frac{\text{TP}}{\text{TP} + \text{FP}}\,,
\end{equation}

\begin{equation}
    \text{Recall} = \frac{\text{TP}}{\text{TP} + \text{FN}}\,,  
\end{equation}

\begin{equation}
   \text{F1}  = \frac{2 \times \text{Precision} \times \text{Recall}}{\text{Precision} + \text{Recall}}\,,
\end{equation}

\begin{equation}
   \text{FPR}  = \frac{\text{FP}}{\text{FP} + \text{TN}}\,,
\end{equation} 
where TP and FN denote the number of true positives and the number of false negatives in the prediction results, respectively. FP and TN are the number of false positives and the number of true negatives. 
We represent TP, FN, FP, and TN in a confusion matrix as shown in \FIG{fig: matrix}.

\begin{figure*}[htbp]
    \centering
    
    \begin{minipage}{0.98\textwidth}
        \centering
        \includegraphics[width=\linewidth]{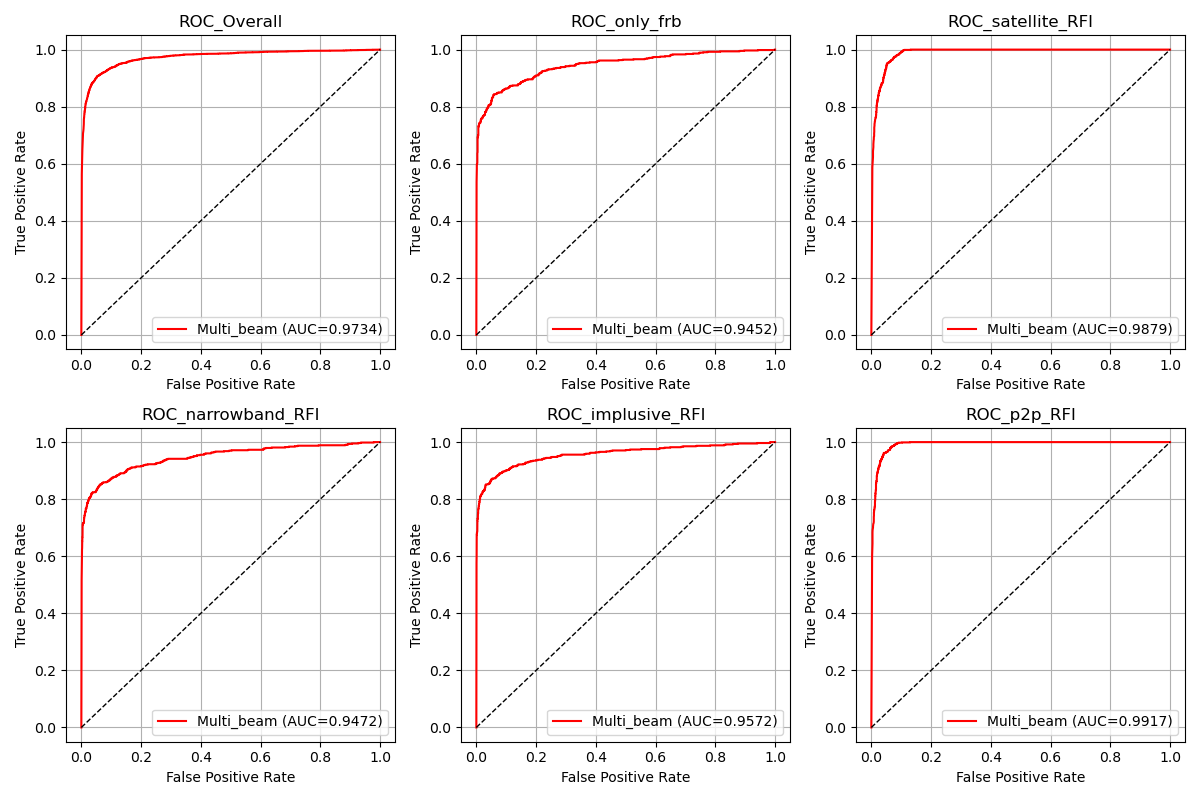}
        
    \end{minipage}\hfill
        
    \caption{Multibeam ROC under different RFI cases. The $ROC\_overall$ plot shows AUC values $>$97\%. The model achieves the AUC exceeding 98.5\% under both satellite and Point-to-Point RFI conditions. The AUC also exceeds 94\% under the other two RFI conditions, indicating that the multibeam model is suitable for FRB searches in the presence of strong RFI. }
    \label{sec:roc}
    
\end{figure*}  

Note that \emph{Accuracy} is the ratio of the correctly predicted data samples among all samples, \emph{Recall} is the ratio of the correctly predicted FRBs among all FRBs in the data, \emph{Precision} is the ratio of the correctly predicted FRBs among all data samples predicted as FRBs. The F1 score is the harmonic mean of precision and recall. It balances the importance of both false positive and false negative, therefore is useful when the FRB data samples and non-FRB data samples are imbalanced.    

FPR measures how many of the negative samples are incorrectly predicted as positive.  

The model correctly identified 3485 positive samples and 3602 negative samples, incorrectly identified 296 positive samples and 298 negative samples. The overall recall and precision of the model are 92.2\% and 92.1\%, respectively.

We further examine the FRB detection performance of the model under different RFI patterns. The recall and precision are shown in Table \ref {sec:Recall table}. The recall with the interferences from the satellite reaches 98.2\%, and 97.3\% with the point-to-point interferences. This demonstrates that our multibeam detection model can successfully detect signals under strong RFIs. 
\begin{table}[htbp]
    \centering
    \caption{The recall and precision in different RFI cases under the multibeam data environment.}
    \label{sec:Recall table}
    \setlength{\tabcolsep}{3.6mm}{\begin{tabular}{ccc}
    
        \hline
        \hline
       
        \textbf{FRBs with RFI types }& \textbf{Recall} &\textbf{Precision}\\ 
        \hline
        
         No RFI & $86.1\%$ & $ 94.6\%$ \\
         Satellites & $98.2\%$ & $ 93.7\%$ \\
         Narrowband & $86.6\%$ & $ 86.1\%$ \\
         Impulsive & $87.5\%$ & $ 90.0\%$ \\
         Point to Point & $97.3\%$ & $ 95.2\%$ \\
      
        \hline
    \end{tabular}}
\end{table}

\subsection{Detection accuracy under different RFI}

We use the receiver operating characteristic (ROC) curve as a metric to evaluate the performance of our multibeam model on FRB search under different RFI patterns. The ROC curve intuitively illustrates the dynamic trade-off between the model's TPR and FPR at different decision thresholds. An ideal ROC curve bends as far as possible toward the upper left corner, indicating that the model achieves high sensitivity while maintaining a low false positive rate, thus demonstrating excellent classification capabilities.  

We calculate the Area Under the Curve (AUC). An AUC value close to 1 indicates excellent classification performance, effectively distinguishing positive from negative samples. As expected, \FIG{sec:roc} clearly demonstrates the excellent performance of our multibeam model, with the ROC curve converging significantly toward the upper left corner and the AUC value confirming its strong discriminative ability.  

In this work, the multibeam model maintains strong discrimination and robustness whether applied to the full sample dataset or under different RFI conditions.

\subsection{Validation with FAST data}

To validate the feasibility of our trained model, we use the actual data from FAST 19-beam surveys, i.e., the Commensal Radio Astronomy FAST Survey (CRAFTS, \citealt{Li+18IMMag,Niu+21ApJ}) and the Galactic Plane Pulsar Snapshot (GPPS, \citealt{Han+21RAA}) survey. Here we basically test the model using five known FRBs detected by the GPPS survey \citep{Zhou+23RAA, Zhou+2023MNRAS}: FRB20230617, FRB20210208, FRB20210705, FRB20210901, and FRB20211005. For these data, the sampling time is 49.152\,$\mu s$, the channel number is 2048. Each data file lasts for either 12.88\,s or 25.77\,s. We downsample the data containing five FRBs to the sizes suitable for the model and use them as a validation set.  

\begin{figure}[htbp]
\centering
\includegraphics[scale=0.6]{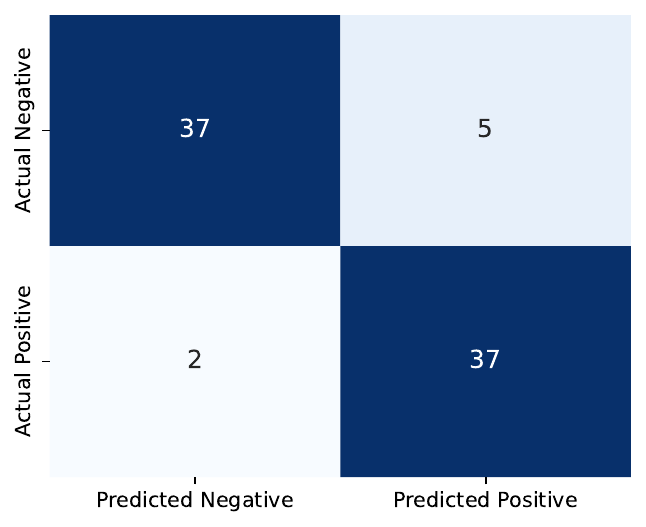}
\caption{Confusion matrices for the validation using actual data. Same as \FIG{fig: matrix}.}
\label{fig: real data}
\end{figure}\hfill

Specifically, the signal in the data file is divided into samples of different durations, for a total of 39 positive samples. These samples are then downsampled to fit the input dimensions of the model. This differs from the simulated data, for which we use skimage's resize function.
We also divide the non-signal data into multiple blocks with the same number of signal samples as the validation dataset. The confusion matrix generated by the model for the real data set can be seen in \FIG{fig: real data}. Some instances may be missed because the signal feature appears in only a very small portion of the sample image. 

\begin{figure}[htbp]
\centering
\begin{minipage}{0.232\textwidth}
    \centering
    \includegraphics[width=\linewidth]{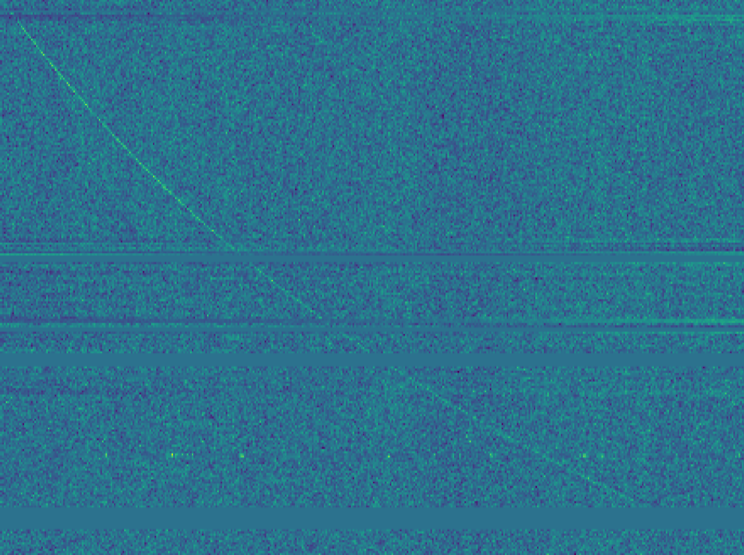}
\end{minipage}\hfill
\begin{minipage}{0.232\textwidth}
    \centering
    \includegraphics[width=\linewidth]{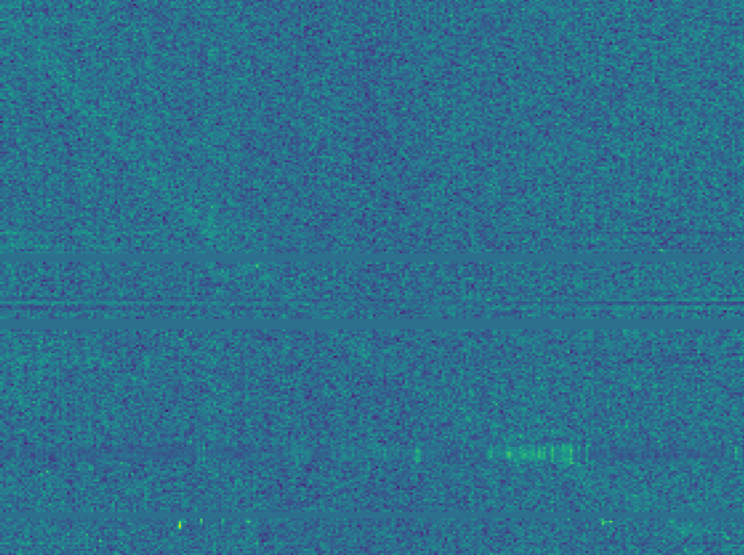}
\end{minipage}\hfill

\caption{Two examples of real FRB signals. In each subplot, the x-axis is the arrival time in units of ms, and the y-axis is the observing frequency from 1000 MHz to 1500 MHz. The real FRB signals have been identified by our EfficientNet model, i.e., 1) FRB 20230617, DM:1219 $\pccm$; 2) FRB 20210901, DM:765 $\pccm$. }
\label{fig:validation_real_data}
\end{figure}

It turns out that our trained EfficientNet model can successfully identify all the actual FRB signals. Two FRBs with prominent SNR difference i.e., FRB 20230617 and FRB 20210901, are detected and shown in \FIG{fig:validation_real_data}. We surprisingly find, while the DMs of real FRBs range from 700 to 2800 $\pccm$, our trained model has been able to recognize all of them out of DM range in our training data. This result may indicate the model is not sensitive with the actual DM values but the sweeping curves in the dynamic spectra. Given that, the DM-free scheme can be flexibly designed for an economic use for future surveys, e.g., detecting an FRB with ultra-high DM.

\hfill

When the model is extended to other telescope system applications, the data parameter space needs to be transformed accordingly, and the model needs to be fine-tuned appropriately.

\section{Discussion} \label{sec:Discussion}

\subsection{Model performance for different significance of signals}

According to the simulation parameters, we define the signal-to-noise ratio (SNR) of mock FRBs as following:
\begin{equation}
\mathrm{SNR} = \frac{S}{S_{\nu,\mathrm{rms}}}\,,
\end{equation}
where $S$ is the peak flux density we set up for a mock FRB, $S_{\nu,\mathrm{rms}}$ is the root-mean-square flux of the simulated system noise, following \EQ{eq:noise}.  

\begin{figure}[htbp]
    \centering
    
    \begin{minipage}{0.48\textwidth}
        \centering
        \includegraphics[width=\linewidth]{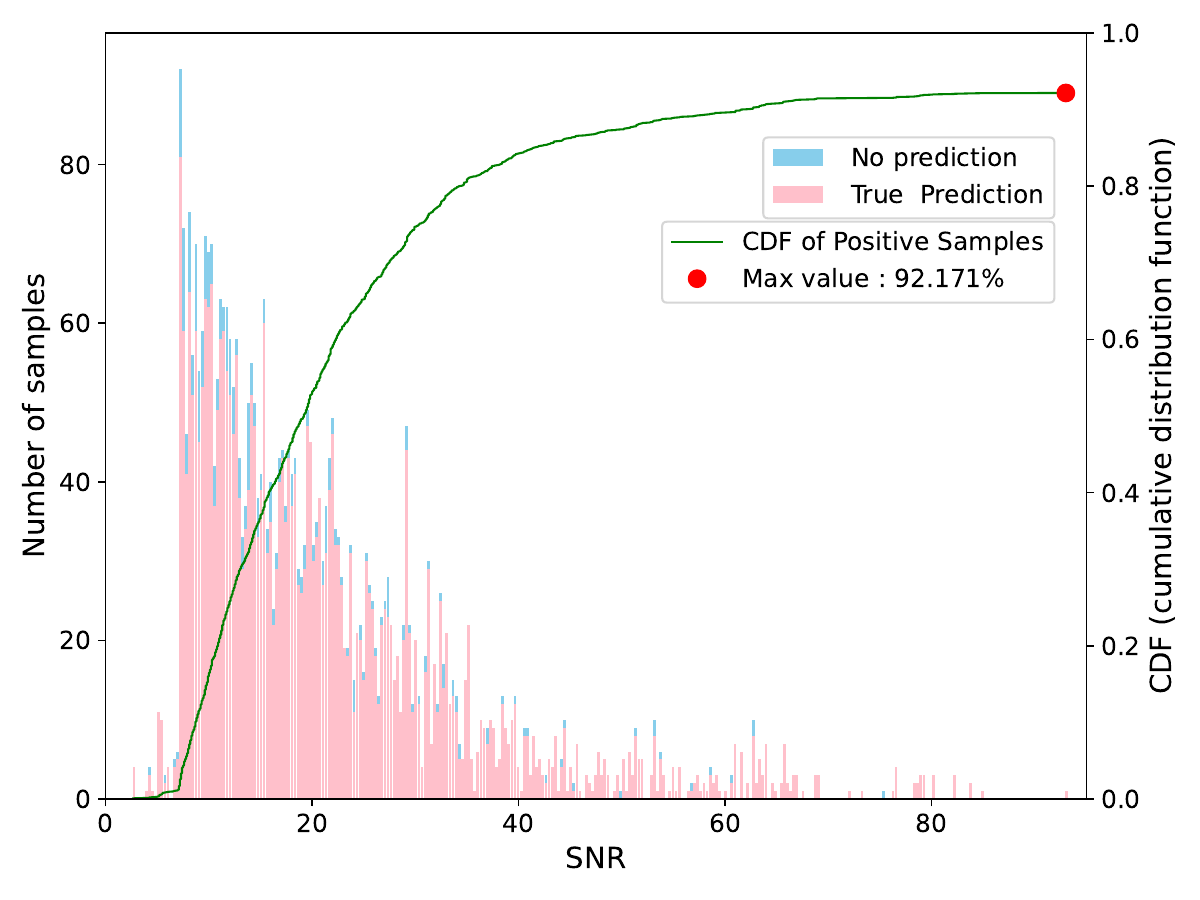}
        
    \end{minipage}\hfill
    
    \caption{A histogram of samples successfully recognized by the model versus all samples. The SNR on the x-axis here represents the SNR of the signal corresponding to each sample, not the SNR of a single sample. The sample unrecognized from the total sample is marked in blue, while the pink represents the sample that has been recognized by the model. The sample ratios are calculated to give a cumulative distribution curve of the proportion of samples recognized by the model to all samples. The maximum point is marked with red circle, and its value is 92.171\%.}
    
   \label{fig:histogram}
\end{figure}

To benchmark the model performance, we count all positive samples as well as the samples detected by the model. Also, we compare the portions of correctly predicted samples along different SNRs, as shown in a histogram (\FIG{fig:histogram}). Note that the same signal has been distributed in different samples, and the signal can be captured by the model in a certain sample. Therefore, some samples with high SNR values are missing in the histogram, but the model can capture other samples of the same signal. 

One can use the cumulative distribution function (CDF) to calculate the cumulative probability corresponding to different SNRs. We thus build CDF of the samples identified by the model and normalize the portion of all positive samples to 1. 
This demonstrates our model is able to identify the most samples even with partial presence of signals, indicating its applicability for real FRB search.

\subsection{Multi-beam vs Single beam}

\begin{figure}[htbp]
\centering
\includegraphics[scale=0.35]{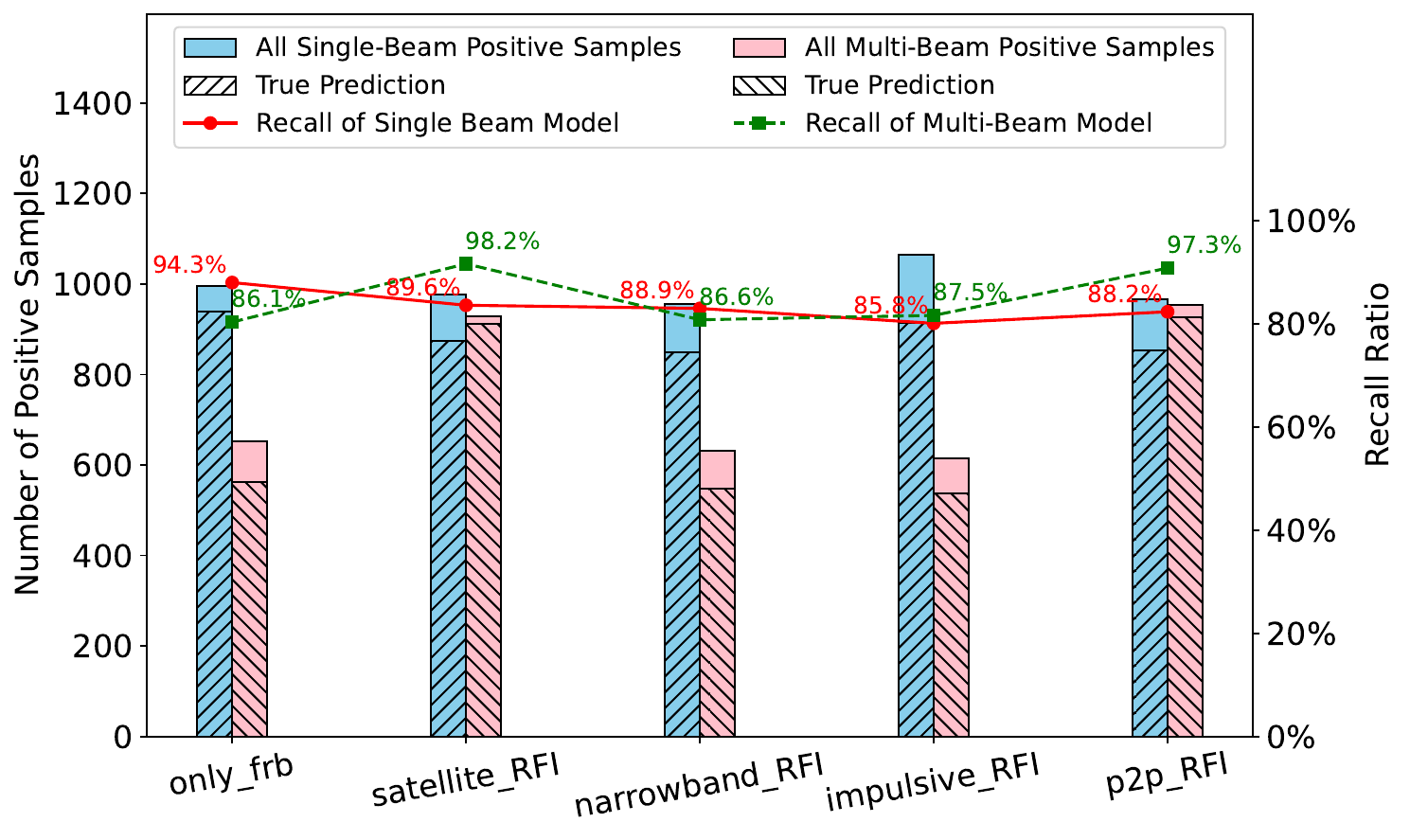}

\caption{A comparison of recall between multibeam and single-beam data. There are 5 cases compared along the x-axis, i.e., 1) only FRBs; 2) satellite RFI; 3) narrowband RFI; 4) impulsive RFI; 5) point-to-point RFI. The dual y-axes denotes number of positive samples and recall ratios, respectively. }
\label{fig: Recall of two models}
\end{figure}\hfill

To verify the advantages of our model, we test the model on single-beam data and compared the results with the multi-beam data model. First, we compare the recall rates of the two models, as shown in the \FIG{fig: Recall of two models}. We present the positive samples identified by the model and calculate the recall. For the data affected by satellite and p2p RFI, the multi-beam model can achieve the recall rates of 98.2\% and 97.3\% respectively, outperforming the single-beam model by nearly 9 percents. If the data are affected by narrowband and impulsive RFI, the difference of the two models becomes less than 2\%. The performance of the single beam is superior to the multi-beam in environments without RFI, but RFI is ubiquitous in real observation.

\begin{figure}[htbp]
\centering
\includegraphics[scale=0.35]{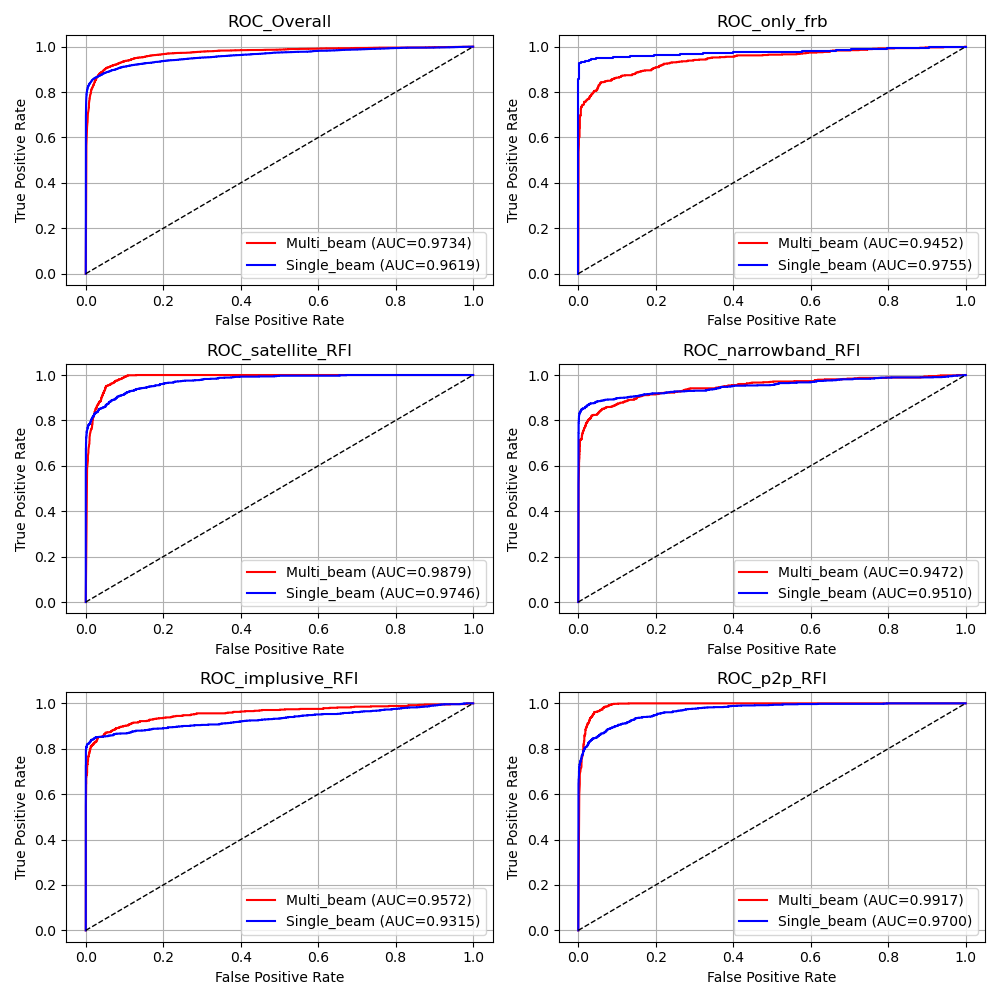}
\caption{Comparison of the ROC curves for the single-beam input and multibeam input of models. The x-axis of each subfigure represents true positives, and the y-axis of each subfigure represents false positives. A higher AUC value, with the ROC curve approaching the upper-left corner, reflects better model distinguishing performance between positive and negative samples. }
\label{fig: roc compare two models}

\end{figure}\hfill

We then compare the ROC curves of the two models in \FIG{fig: roc compare two models}. The figure shows that when the data is affected by satellite, p2p, and impulsive RFI, the model performs significantly better on multi-beam data than on single-beam data. Specifically, the multi-beam model achieves an AUC (area under the curve) of 98.79\% for satellite RFI, which is 1.33\% higher than that of the single-beam model. For p2p RFI, the AUC reaches 99.17\%, corresponding to an improvement of 2.17\%. For impulsive RFI, the multi-beam model attains an AUC of 95.72\%, exceeding the single-beam model by 2.57\%. Overall, the multi-beam model consistently outperforms the single-beam model, yielding an average AUC improvement of approximately 1.2\%.

\subsection{Real-time searching compared with CPU-based algorithms}

\begin{figure}[ht]
\centering
\includegraphics[scale=0.35]{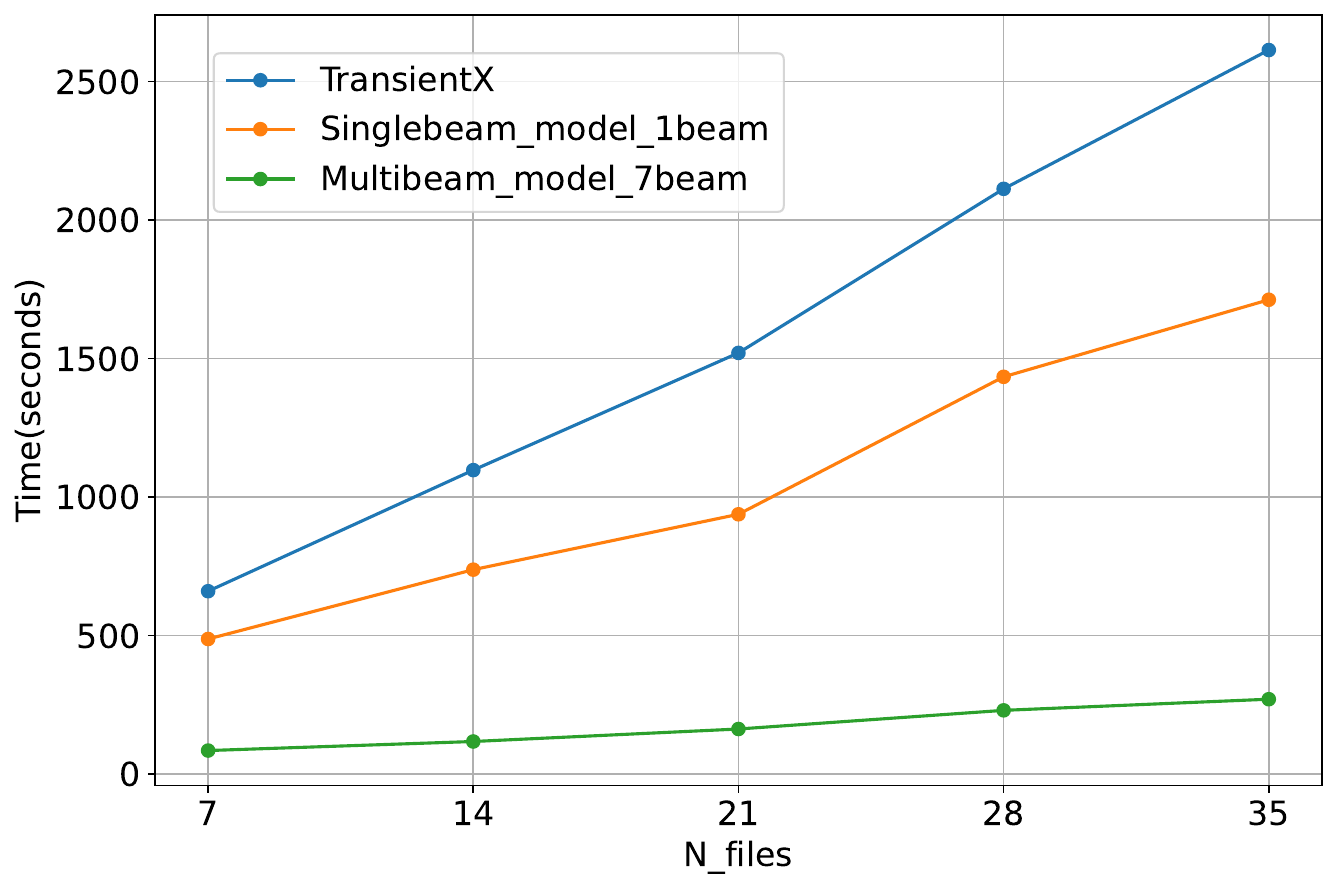}
\caption{Comparison of computing time between our model and TransientX. The x-axis is the number of FRB data files used for this test, and the y-axis is the time consumed during computing. }
\label{fig: Efficiency}
\end{figure}

We benchmark the computing speed of our model by comparing it with TransientX\footnote{\href{https://github.com/ypmen/TransientX}{https://github.com/ypmen/TransientX}}\citep{Men+24AA}, which is a CPU-based FRB searching software. We use 7 beam files (including the signal beam file and six surrounding beam files) from one to five signal sources, for a total of 35 files. We compare the fastest performance of the multi-beam model, the single-beam model, and the single-threaded TransientX run time, as shown in \FIG{fig: Efficiency}. By increasing data amount, we find our model is able to reduce the data 9 times faster than TransientX approximately, as it can recognize the multibeam data simultaneously and report all the results together. 

Here we consider 7-beam case for a FAST-like telescope. The requested data is sampled every 98.304 $\mu$s, digitized by 8-bit with 4096 channels in a frequency range from 1--1.5 GHz. If full polarization information (4 channels) is recorded, the data rate can reach 3.8 TB/hr estimated with the following equation:
\begin{equation}
    \eta=(3600/t_\mathrm{samp})\times N_\mathrm{bit}\times N_\mathrm{pol}\times N_\mathrm{chan}\times N_\mathrm{beam}\,.
\end{equation}

In our case, the optimized Efficient model can process 35 data files in 270 seconds using a single GPU (71-GB memory) with a throughput of 0.92 TB/hr. In practice, most blind searches use 1- or 2-bits to digitalize data, which further improve the detection throughput. Our method has a potential to achieve real-time FRB search with GPUs. 

As TransientX is implemented using CPUs, one may argue that our GPU-based method incurs a higher cost for the performance gain. However, the Intel Xeon Platinum 8358P processor (128 cores, 2.6 GHz) we run TransientX on actually costs more than the NVIDIA GeForce RTX 4090 (24GB) graphics cards (24 GB) we run our machine learning models in Figure ~\ref{fig: Efficiency}, which shows the cost-effectiveness of our method. 

\hfill

\subsection{DM estimation without dedispersion}

Our method is able to classify the continuous chunks of multibeam data as potential FRBs or non-FRBs without applying dedispersion on the data. This eliminates the computation cost of dedispersing a large number of non-FRB data chunks. However, the actual DM remains unknown for data chunks classified as FRBs, which may require dedispersion computing. This can still be a burden when the DM search range is large and the number of candidate FRB data chunks is high. We further give a method to directly estimate the DM value using a pretrained Variational Autoencoder, or VAE~\citep{kingma2019introduction} for a candidate FRB data chunk as below.

The VAE consists of (i) an \emph{encoder} $q_\phi(\mathbf{z}\mid \mathbf{x})$ that maps the input data chunk $\mathbf{x}$ to a Gaussian posterior over a latent variable $\mathbf{z}\in\mathbb{R}^d$ and $p(\mathbf{z})=\mathcal{N}(\mathbf{z}; \mathbf{0},\mathbf{I})$, (ii) a \emph{decoder} $p_\theta(\mathbf{x}\mid \mathbf{z})$ that reconstructs the FRB signal in $\mathbf{x}$, denoted by $\mathbf{x_{sig}}$ and the estimated DM values $\mathbf{\hat{y}}$.

\begin{figure}[ht]
	\centering
	\includegraphics[width=.45\textwidth]{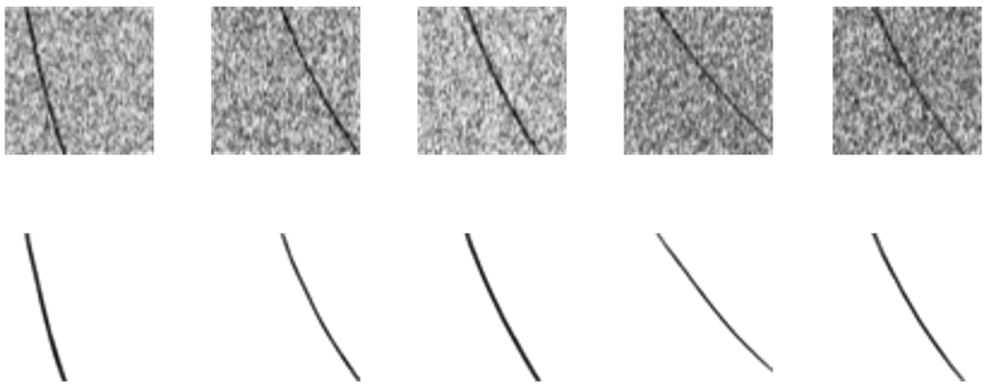}
	\caption{Examples of DM estimation using VAE. The top row shows the data chunks with FRB and the bottom row shows the reconstructed FRB signals. The true DM and estimated DM pairs for each column are as follows: (141, 157), (287, 274), (272, 270), (467, 469), (326, 372)}
	\label{fig:dm-estimates}
\end{figure}

To train the model, we first use our simulator to  generate the FRB signals $\mathbf{x_{sig}}$ with given DM values, denoted by $\mathbf{y}$, and generate the corresponding data chunks containing the signals, i.e., $\mathbf{x}$. We then feed $(\mathbf{x}, \mathbf{x_{sig}}, \mathbf{y})$ to the VAE. Through $q_\phi(\mathbf{z}\mid \mathbf{x})$, we obtain $\mathbf{\hat{y}} = \mathbb{E}_{p(z)}[z]$. The loss of $p_\theta(\mathbf{x}\mid \mathbf{z})$ is the combination of the cross entropy loss between $\mathbf{x_{sig}}$ and the reconstructed $\mathbf{x}$, and the mean square error (MSE) between $\mathbf{y}$ and $\mathbf{\hat{y}}$. By minimizing the difference between the mean of the latent variable and the DM of the known FRB, the VAE learns the dispersion patterns in FRB data chunks. This approach can therefore further reduce the computational cost of dedispersion. Example results of this approach are shown in \FIG{fig:dm-estimates}.

\subsection{Potential for future all-sky surveys}
Different from the FAST telescope with only 19 beams currently, next-generation radio facilities, such as Square Kilometre Array (SKA), will be able to generate a huge number of coherent and incoherent beams to cover the entire southern sky \citep{Sokolowski+21PASA}. Thanks to the upgrade on telescope receiving systems, the concept of ``All-Sky'' has been shaped as a result of some emerging instruments, e.g., dipole-antenna stations \citep{Connor+21PASP} or array \citep{Lin+22PASP}, ground-based phased array \citep{Luo+24PASA}. The huge data rate yielded by thousands of beams will absolutely challenge real-time blind search for fast radio transients. Due to a large amount of time consumed for dedispersion by conventional approaches, one can hardly generate and sift FRB candidates from all the beamforming data in real time. Moreover, it is impossible to deal with that issue in the offline scenario because of limited data storage. Hence, a DM-free searching pipeline for multibeam data is not optional but necessary scheme for the forthcoming facilities.

We preliminarily demonstrate that it is feasible to deploy a DM-free ML framework for multibeam data. After training based on the EfficientNet model, we test and validate the results for both simulation and actual data. The precision-recall performance is optimistic, and we can see the clear difference in the computing efficiencies between traditional and ML algorithms. Since this is a simple attempt for limited multibeam data, we pretty much look forward to promoting more ML-based transient searching applications on massive data in the near future.

\section{Conclusions} \label{sec:Conclusions}

In this paper, we propose a novel DM-free FRB search method based on deep neural network. By training EfficientNet-based models to analyze the dynamic spectra from multiple beams simultaneously, our approach reduces the cost of DM trials in traditional algorithms.

Our results show the proposed method achieves high FRB detection performance on simulated test data, with both precision and recall around 92\%. The method maintains robust sensitivity even in the presence of strong radio frequency interference. In particular, when multiple beams are used as input, the network naturally learns to suppress terrestrial RFI (which appears in all beams) while highlighting an astrophysical burst confined to a single beam. This led to excellent recall rates (exceeding 95\% in our tests) for FRB signals under various RFI conditions, substantially outperforming a single-beam approach. 

We also apply the trained model on real FAST data containing known FRBs. The model can reliably identify bursts with $\gtrsim 95\%$ recall and a low false positive rate, indicating our model has good generalization capability on real survey data.

In addition to its accuracy, a major advantage of the ML pipeline is its computational efficiency. By eliminating brute-force dedispersion and using fast GPU inference, our DM-free search runs faster than conventional single-pulse search software (e.g., \textsc{presto}, \textsc{TransientX}), which must test thousands of DM trials. In practical terms, our approach can process high-volume data streams in a fraction of the time taken by those traditional algorithms. 

In summary, our DM-free, multibeam machine learning approach provides an efficient and reliable new paradigm for FRB searches. This speed-up, combined with the model’s RFI resilience, paves the way for real-time FRB discovery in large-scale surveys. This paper is the first part of a series on DM-free FRB search; in a forthcoming work we will attempt to upgrade the current ML/AI approach for actual surveys in the real time by measuring the DM values of FRB candidates accurately without dedispersion.

\section*{Acknowledgments}
This work makes use of the GPU cluster at Guangzhou University. Y.C., R.L. and S.Q.Z. are supported by the National Natural Science Foundation of China (NSFC) Grant No.\,12303042. Y.K.Z is supported by the Postdoctoral Fellowship Program and China Postdoctoral Science Foundation under Grant Number BX20250158.
We thank Xiaoyun Ma, Bing Li, and Jiguang Lu for helpful suggestions.

\section*{Data and Code Availability }

The source code used in this work is publicly available on GitHub at \url{https://github.com/cyyd12356-creator/multibeam-frb-search}. A subset of the test datasets used for evaluation is available on Hugging Face at \url{https://huggingface.co/datasets/cyyd12356/data}. The trained models presented in this study are also publicly available on Hugging Face at \url{https://huggingface.co/cyyd12356/model}.  

We are pleased to provide further data products, trained models, and implementation details to interested readers upon reasonable request, in order to facilitate reproducibility and future research.

\bibliography{refs}{}
\bibliographystyle{aasjournal}

\end{document}